\documentclass[traditabstract]{aa}
\usepackage{aas_macros}
\usepackage{hyperref}
\hypersetup{colorlinks=true,citecolor=blue}

\usepackage{booktabs,adjustbox}
\usepackage{longtable}
\usepackage{array}
\usepackage{url}
\usepackage{algorithmicx}
\usepackage{algorithm}% http://ctan.org/pkg/algorithms
\usepackage{algpseudocode}
\usepackage{scrextend}
\usepackage{multirow}
\usepackage{xcolor}

\usepackage{amsmath}
\usepackage{appendix}
\usepackage{rotating}
\usepackage{float}
\usepackage{graphicx}	
\usepackage{txfonts}
\usepackage{natbib} 
\usepackage{xspace}
\usepackage{url}
\bibpunct{(}{)}{;}{a}{}{,} % to follow the A&A style
\usepackage{subfigure}

\newcolumntype{x}[1]{>{\centering\arraybackslash\hspace{0pt}}p{#1}}

\begin{document}
\title{Multi-band morpho-Spectral Component Analysis Deblending Tool (MuSCADeT):  Deblending colourful objects}
\titlerunning{MCA-based lens-source separator}

\author{R. Joseph\inst{1} \and F. Courbin\inst{1}  \and J.-L. Starck\inst{2}
}
\authorrunning{R. Joseph et al.}

\institute{
Laboratoire d'astrophysique, Ecole Polytechnique F\'ed\'erale de Lausanne (EPFL), Observatoire de Sauverny, CH-1290 Versoix, Switzerland \and  
Laboratoire AIM, CEA/DSM-CNRS-Universite Paris Diderot, Irfu, Service d'Astrophysique, CEA Saclay, Orme des Merisiers, 91191 Gif-sur-Yvette, France
}

\date{Received ; accepted }

\abstract{We introduce a new algorithm for colour separation and deblending of multi-band astronomical images called {\tt MuSCADeT}  which is based on Morpho-spectral Component Analysis of multi-band images.  The {\tt MuSCADeT} algorithm takes advantage of the sparsity of astronomical objects in morphological dictionaries such as wavelets  and their differences in spectral energy distribution (SED)  across multi-band observations. This allows us to devise a  model independent and automated  approach to separate objects with different colours. We show with simulations that we are able to separate highly blended objects and that our algorithm is robust  against SED variations of objects across the field of view. To confront our algorithm  with real data, we use HST images of the strong lensing galaxy cluster MACS~J1149+2223 and we show that {\tt MuSCADeT} performs better than traditional profile-fitting techniques in deblending the foreground lensing galaxies from background lensed galaxies. Although the main driver for our work is the deblending of strong gravitational lenses, our method is fit to be used for any purpose related to deblending of objects in astronomical images. An example of such an application is the separation of the red and blue stellar populations of a spiral galaxy in the galaxy cluster Abell~2744. We provide a python package along with all simulations and routines used in this paper to contribute to reproducible research efforts. Codes can be found at \url{http://lastro.epfl.ch/page-126973.html}.}

\keywords{Methods: data analysis -- Gravitational lensing: strong -- Galaxies: surveys --  stellar populations}

\maketitle

\section{Introduction}
\label{sec:Intro}

	Astronomical objects are often seen merged or blended on the plane of the sky. This blending can be apparent, because objects at different distances are seen in projection on the plane of the sky or, real, because different objects at the same distance are physically overlapping. 

	Whatever the reason for the blending, reliable deblending techniques are mandatory for astrophysical projects to meet their scientific objectives. Among the many possible examples, blends of galaxies of different colours can impact performances of photometric redshift algorithms \citep[e.g.][]{Perez2010, Bellagamba2012, Parker2012, Hsu2014} and conclusions of stellar populations stu\-dies \citep[e.g.][]{Yan2014}. Obviously, blending also affects the determination of morphological properties of astronomical objects, for example the shape measurement of faint galaxies in  weak lensing cosmological surveys \citep[e.g.][]{Chang2013, Arneson2013}. In strong gravitational lensing, deblending of the foreground len\-sing object from the background lensed sources is essential, for example as shown at galaxy scale by \citet[][]{Gavazzi2007} and at cluster-scale by \citet{Massey2015}. This is true for at least two reasons. First, one needs to map the visible mass in the lensing object precisely, either to use it as a prior to guide the lens modelling or to infer the mass-to-light ratio in the lens. Second, the image of lensed source must be isolated in the best possible way. Any faint extended arc-like structure, clump, or star-forming region must be seen precisely with minimum light contamination from the lensing object. Our ability to constrain the mass model is completely driven by the amount of details seen in the lensed source, which represent as many observational constraints. 

	Many of the current techniques to deblend astronomical objects are limi\-ted to analytical modelling of their light distribution either in single band \citep[e.g. {\tt PSFex},][]{Bertin2011} or multi bands, sometimes including a simultaneous fit of many overlapping objects \citep[{\tt Megamorph};][]{Vika2013, Vika2015}. Alternatively, some me\-thods make use of high resolution images to flag blended objects and then measure them at different wavelengths using images of lower spatial resolution \citep[e.g.][]{Laidler2007}. Popular softwares like {\tt Sextractor} \citep{Bertin1996} use image segmentation to separate blends, which is a technique that was further improved by \citet{Zheng2015}. Other techniques include machine learning, recently used in the area of strong gravitational lensing to subtract the light of bright galaxies and to unveil possible lensed background objects without invoking any analytical representation of the galaxies to subtract. This technique is based on a principal component decomposition of the galaxy images using large samples of single-band imaging data \citep{Joseph2014}. A step forwards is to use multi-band images to separate the objects in the lens plane and source plane, also using the colour information. Recent methods have started to make use of multi-band information and combine source position from high resolution images with profile fitting to deblend lower resolution bands \citep{Merlin2015}. Another example is given in \cite{Hocking2015} where neural networks are used to identify objects with different colours in multi-band images. 
		
With the current burst of wide-field sky surveys (DES, KIDS, HSC, Euclid, LSST and WFIRST), data in many optical and near-IR bands will become available. The present paper describes a technique taking advantage of these multi-band data to address the deblending problem using both colour information and a spatial prior, but not involving any analytical modelling of the light distribution of the lensed and lensing objects.  Our work is based on a multi-channel extension of the morphological component analysis (MCA) presented in \citet{Starck2004}. We illustrate the performances of the algorithm with numerical experiments and with real HST data of strong gravitational lenses. 

%	In the last decade, several similar techniques have emerged that are more adapted to solve other specific source separation problems. For instance, Multi-channel Morphological Component Analysis (MMCA) \citep{Bobin2006} solves the problem of source separation in multi-channel data assuming the components have different morphologies and thus can be represented in different dictionnaries;  Generalised Morphological Component Analysis (GMCA) \citep{Bobin2007b, Bobin2008} addresses the problem of blind source separation of morphologically distinct sources; Local-GMCA (L-GMCA) \citep{Bobin2013,Bobin2011} uses the principles of GMCA on adaptive patches of the sky and locally estimates the mixing matrix, it was successfully applied to CMB the reconstruction problem in \cite{Bobin2014};  non-negative GMAC (nGMCA) \citep{Rapin2013,Rapin2014} accounts for the positivity of the observables in direct space along with the sparsity of the sources in transformed space to perform blind source separation; Adaptive MCA (AMCA) \citep{Bobin2015} aims at solving source separation problems for partially correlated components.

This paper is organised as follows: In section \ref{sect:model}, we introduce the mathematical model we use to understand and separate objects with different colours. In section \ref{sect:MCA}, we describe the mathematical technique used to solve the problem of colour separation in our approach, that is to say morphological component analysis. In section \ref{sec:MuSCADeT}, we detail our implementation of the MuSCADeT algorithm. Section \ref{sec:Test} shows the performance of our algorithm on simulations that test realistic problems encountered in deblending. We apply our method to real astronomical images from the Hubble Space Telescope in section \ref{sec:Data} with the galaxy clusters MACS J1149+2223 and Abell~2744. We compare our results with current model fitting methods. Section \ref{sec:repro} provides useful information to reproduce our results from the code we made freely available.

% ====================

\section{Deblending and strong lensing}
\label{sect:model}
\subsection{The source separation problem}
\label{sec:Source}
We assume the observed data $\{y_i\}_{i=1,.., N_b}$ in the band $ i $  can be represented as
\begin{equation}
y_{i} [k] =   \sum_{j=1}^{N_o} s_{i,j}  o_{j} [k]  + z_{i} [k],
\label{eq_model_strong}
\end{equation}
where $o_j$ are the different observed sources, $s_{i,j}$ is the contribution of the j-th source in the observation $y_i$,  
$N_o$ the number of sources, $N_b$ the number of bands, $z_{i}$ is an additive Gaussian noise,   $k$ is the pixel index ($k = 1 ... N_p$), 
and $N_p$ is the number of pixels.

The parameter $s_{*,j}$ corresponds to the spectral energy distribution (SED) of the source $o_{j}$.
The deblending problem consists in finding the different objects $o_j$, which is somewhat complicated since their SED are not known and
even the number of objects is not known.

However, several galaxies may have similar colour properties and, therefore, share the same SED, so we can simplify Eq.~\ref{eq_model_strong} by considering
the data containing only $N_s$  groups of sources, such as, for instance  early- and late-type galaxies, 
and we can restrict the deblending problem to only extract  these two groups.
We note $x_j$ ($j = 1.. N_s$) the image which contains the sum of all objects belonging to  the group 
$j$, i.e. $x_j [k] =  \sum_{l=1}^{N_{o}^{(j)}}  o_{l} [k] $\textbf,
where $N_o^{(j)}$ is the number of sources in the group $j$. We can write
\begin{equation}
y_{i} [k] =  \sum_{j=1}^{N_s} a_{i,j}  x_{j} [k]  + z_{i} [k].
\label{eq_linear_model_strong}
\end{equation}
Even if this equation looks very similar to Eq.~\ref{eq_model_strong}, it is in fact  simpler since $N_s$ is smaller than $N_o$.
As a given component $x_j$ contains several astrophysical sources, it also gives us more statistics to derive its SED.
This linear mixture model can be recast in the following matrix form
\begin{equation}
Y = A X + Z,
 \label{eq:Source}
\end{equation}
where $Y$ is a $N_b \times N_p$ matrix, $A$ is the SED mixing matrix, and $X$ is the $N_s \times N_p$ matrix, which contains 
the components $x_j$. 

\begin{center}
\begin{figure}[t!]
\includegraphics[scale=0.27]{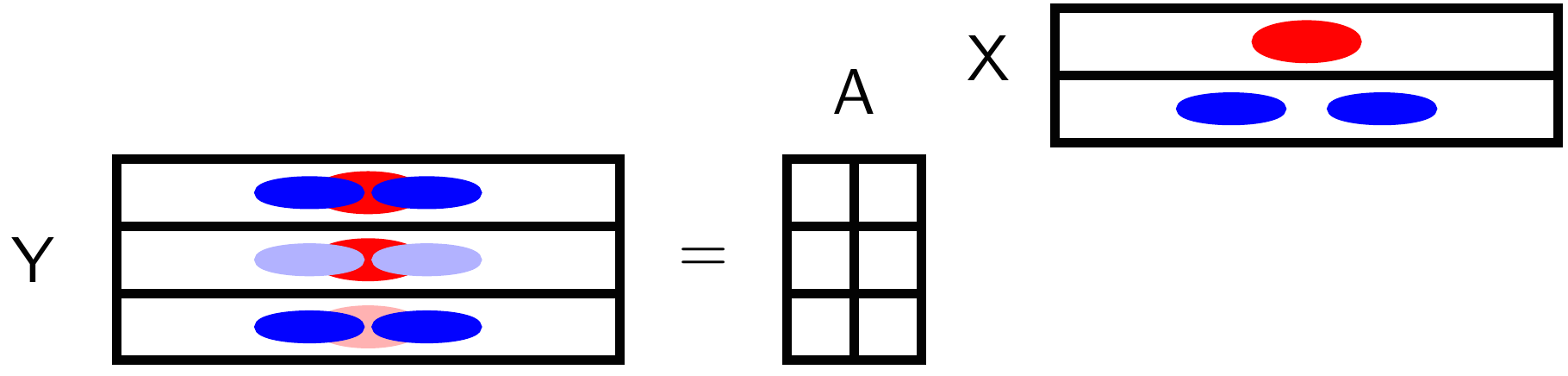}
\caption{ \label{fig:ASX} Illustration of the blind source separation in the case of two sources. To make the figure simple, the images in each band are represented as lines in the $Y$ matrix. Sources are lines in the $X$ matrix. On the sketch we figure a red object in the first source and two blue objects in the second source. Matrix $A$ contains the mixing coefficients that allow various combinations of elements of $X$ to produce $Y$.}
\end{figure}
\end{center}

To sum up,  we consider that each band is a weighted mix of $n_s$ colour components. 
In the statistical  literature, each component is called a source (and it should not be mixed with an astrophysical source). 
This general problem is called a blind source separation problem (BSS), i.e. estimating both $A$ and $X$ knowing only $Y$.
The weight for a given source is the value of the associated SED at the corresponding wavelength. 
Figure ~\ref{fig:ASX} illustrates the BSS in the case of two sources relative to two populations of galaxies (red and blue galaxies in the figure).

%{\bf Le dessin n'est pas correct, car une bande est reprÃ©sentÃ©e par une ligne dans la matrice Y, idem pour X.  Il faut changer aussi les notations. \textcolor{red}{Je vois pas le problÃšme}}

\subsection{Determination of the mixing matrix A}
\label{sec:PCA}

	The $A$ matrix is central to modelling of multi-band data as it describes the contribution of the different sources to the images taken at different wavelengths. In practice, the elements of $A$ are the SEDs of the objects in the sources $X$. They can be assumed, for example as template spectra for objects of a given type, or they can be measured because spectroscopic data are available for at least some of the objects in the field of view. In most cases, however, the matrix $A$ needs to be estimated solely from the multi-band imaging data, $Y$. In order to do this, we use a method based on a principal component analysis \citep[PCA;][]{Jolliffe1986} of the data. 
	
We consider the multi-band data as an ensemble of vectors $\{y_{i=1..N_b}[k]\}$, where the pixel values in band $i$ at the spatial  location $k$ are stored in $y_i[k]$, as previously. In other words, $\{y_{i=1..N_b}[k]\}$ is the measured SED at location k. 

\begin{figure*}[t]
\begin{tabular}{lll}
\hspace*{-2mm}
\includegraphics[width = 5.2cm, height = 5.2cm]{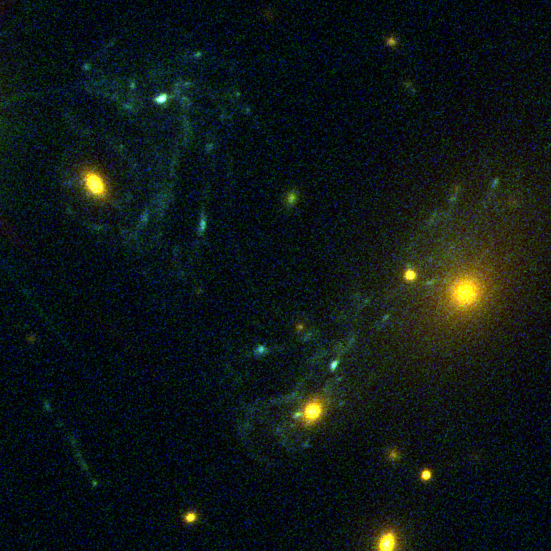} &
\hskip -5pt
\includegraphics[scale=0.35]{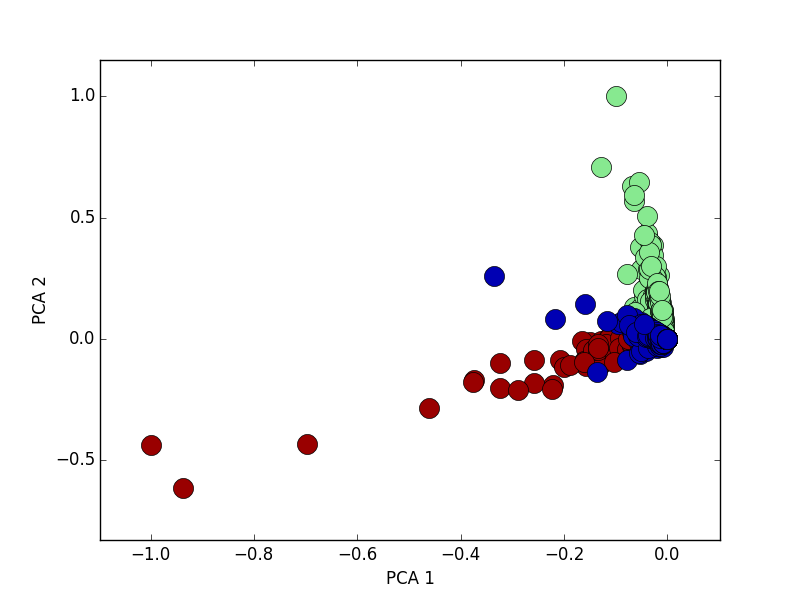} &
\hskip -5pt
\includegraphics[width = 5.2cm, height = 5.2cm, clip = true]{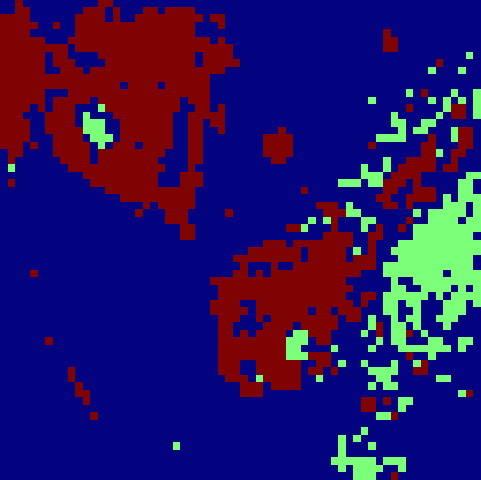} 
\end{tabular}
\caption{Illustration of the PCA colour selection. {\it Left:} HST image of the "Refsdal lens" in the galaxy cluster MACS~J1149.6+2223. {\it Middle:} distribution of the first two PCA coefficients. The red and green dots correspond to the coefficients attributed to the first and second sources by {\tt MuSCADeT}, respectively. Blue dots are rejected coefficients. {\it Right:} corresponding spatial distribution of colours as detected via PCA.} \label{fig:PCA}
\end{figure*}

Following our definition of a source, two pixels belonging to a given source $x$ have proportional SEDs. To build the mixing matrix $A$, one solution is to preselect obvious objects belonging to a same source and to average their SEDs, thus approximating the mixing coefficients corresponding to their source. A more subtle way to do this is to perform a principal component analysis of all SEDs belonging to bright objects and to look for proportional vectors. The details of this procedure can be summarised as follows:
\begin{itemize}

\item We select the brightest objects in all bands and perform the PCA of the SEDs at pixel locations with high signal to noise. In practice, this is done by applying a wavelet filtering of all bands. Also, to save computation time, we rebin images to $64\times64$ pixels in size.
			
\item We perform a clustering analysis of the first two PCA components: the linearity and orthogonality of the PCA decomposition implies that proportional vectors see their respective PCA coefficients distributed along the same hyperplane in the PCA space. In other words, vectors with proportional SEDs have their first two components, PC1 and PC2, distributed along lines in the PC1-PC2 space, as illustrated in Fig.~\ref{fig:PCA},
		
\item We identify the SEDs that have proportional PC1 and PC2 coefficients and average them. Coefficients that are judged too faint or too ambiguous (i.e. they could be a mix of both sources) are rejected,

\item We store the resulting mean SEDs as a column in the $A$ matrix.
		
\end{itemize}

This algorithm shows good results in identifying objects with different dominant colours (Fig.~\ref{fig:PCA}), but its capabilities in terms of deblending are rather limited when distinct sources spatially overlap. The above PCA analysis is only a spectral analysis. The MCA method proposed in this paper combines the strengths of the morphological analysis and spectral analysis to design a reliable deblending algorithm. 

% ====================

\section{Morphological component analysis}
\label{sect:MCA}
The morphological component analysis (MCA) method \citep{Starck2004,starck:book10} allows us to separate 
several components in a given image based on their morphological diversity. 
Indeed, it was shown that it  is possible to disentangle two (or more) signals mixed into one observable, this is only based on the fact that each of those signals can be sparsely represented in their respective data representation domains, called dictionaries, but not in the other's.  For instance, one could separate a periodic signal from a Gaussian profile in an image based on Fourier transform (associated with the periodic signal) and the wavelet transform (for the Gaussian profile). The projection of the mixing over the Fourier dictionary shows enhanced contribution from the periodic component whereas wavelet space shows higher coefficients for the Gaussian component.
 
\subsection{Separation using a sparsity prior}
MCA is based on the concept of sparse signal representation.
A signal is sparse in a dictionary  $\Phi$ when it can be well represented,  such that $x = \Phi \alpha = \sum_i \phi_i \alpha_i$ and only few coefficients $\alpha$ 
are different from zero.  Some dictionaries, such as Fourier or wavelets, have implicit fast transformation and reconstruction operators that allows us to derive the coefficients $\alpha$ from $x$ (and also to derive $x$ from $\alpha$) efficiently without having the elements of matrix $\Phi$ in memory.  In inverse problems,  a sparse solution is imposed by adding an $\ell_0$-norm penalisation term to the data fidelity attachment term.
MCA is an iterative algorithm, which separates a single image $Y$ into $J$ components $x_j$, by solving 
\begin{equation}
\min_{x_1, ..., x_J}  \sum_{j=1}^J \parallel \Phi^* x_j \parallel _0   ~~ s.t. ~~ \parallel  Y - \sum_{j=1}^J   x_j \parallel^2 \le \sigma,
\label{eq_mca}
\end{equation} 
where $\Phi^*x = \alpha$ and $\sigma$ is the noise standard deviation of the noise.
Full details can be found in \citet{starck:book10}.
This method has been used  to extract   filamentary clouds in Herschel data \citep{andre2010}
or, more recently, to improve SNIa detection in the SuperNova Legacy Survey data set \citep{moller2015}.

\subsection{Multi-band dictionaries}
As we have multi-band data, we need to use morpho-spectral diversity. The dictionary $\Phi_i$ related to a given component $x_i$ is therefore a tensorial product of a spectral dictionary ${\cal S}_j$ with the spatial dictionary $\Psi_j$, i.e. $\Phi_j = {\cal S}_j \Psi_j$

\subsubsection{Spatial dictionary}
In the case of strong lensing, the diversity between the components is mainly related  to a different spectral morphology. 
Therefore we can reasonably use the same spatial dictionary, and in this strong lensing application we use 
 the starlet dictionary \citep{Starck2007}. Starlet transform is an isotropic, undecimated wavelet transform that is computed  using consecutive convolutions by a B$-$spline profile as a scaling function \citep{Starck2007a}. The resulting starlet representation is an overcomplete set of coefficients that represent variations in an image at different scales, and is particularly suited to represent astronomical images \citep{Starck2006}. 

\subsubsection{Morpho-spectral dictionary}
We note $\Psi$ the starlet dictionary, we have $x_j = \Psi \alpha_j$ where $\alpha_j $ are the starlet coefficients of the $j$th source. 
A good choice for the morpho-spectral dictionary is to take $\Phi_j = a_j \Psi $,  where $a_j$ is the $j$th column of the matrix $A$.
The data attachment  term  for multichannel data can be written as
\begin{eqnarray}
L  & = &  \parallel  Y - \sum_{j=1}^J   \Phi_j  \alpha_j  \parallel^2 =  \parallel  Y - \sum_{j=1}^J   a_j \Psi  \alpha_j  \parallel^2  \nonumber \\
     & = & \parallel  Y - \sum_{j=1}^J   a_j x_j  \parallel^2 =  \parallel  Y -  A X   \parallel^2.
\end{eqnarray} 
 The multichannel MCA hence consists in changing the data attachment term only, and we need to solve
 \begin{equation}
\min_{X}  \sum_{j=1}^J \parallel \Psi^* x_j \parallel _0   ~~ s.t. ~~ \parallel  Y -  A X   \parallel^2   \le \sigma.
\label{eq_mmca}
\end{equation} 

\subsubsection{Lagrangian form and positivity}
The sparse recovery problem can be formulated under an augmented Lagrangian form 
\begin{equation}
\min_{X} \parallel  Y -  A X   \parallel^2 +  \sum_{j=1}^J \lambda_j \parallel \Psi^* x_j \parallel _0   
\label{eq_lag_mmca}
\end{equation} 
and we can add a positivity constraint to the solution, so we need to solve
\begin{equation}
\min_{X} \parallel  Y -  A X   \parallel^2 +  \sum_{j=1}^J \lambda_j \parallel \Psi^* x_j \parallel _0   ~~ s.t. ~ \forall j, ~   x_j \ge 0,
\label{eq_lag_mmca_pos}
\end{equation} 
	
where $\lambda_j$ accounts for the sparsity of each component $x_j$ in its own morpho-spectral dictionary. The next section describes how this equation can be solved.

%====================================

\section{MuSCADeT algorithm}
\label{sec:MuSCADeT}

The strong lens sparse deblending iterative algorithm ({\tt MuSCADeT}) is an extension of the MCA algorithm,
consisting in applying at each iteration three main steps:
\begin{enumerate}
\item Perform a gradient step: $U = X^{(n)} + \mu A^t (Y - A X^{(n)})$.
\item Solve for each $j$: ${min}_{x^{(n+1)}_j}  \parallel u_j - x^{(n+1)}_j || + \lambda_j  \parallel  \Psi^* x^{(n+1)}_j \parallel _0$, and 
set to zero negative entries in $x^{(n+1)}_j$.
\item Decrease $\lambda_j$.
\end{enumerate}

	In this algorithm $\mu$ is the gradient step, derived from the 2-norm of matrix $A$ \citep{Higham1992} such that $\mu = 2/||A||_2$. 

	This algorithm is also directly related to the proximal forward-backward algorithm \citep{Combettes2005}. A very nice aspect of this 
algorithm is that the minimisation involved in the second step do not require any iteration, and is obtained by 
$x^{(n+1)}_j = \Delta_{\Psi, \lambda_j} (u_j)$, where $\Delta_{\Psi, \lambda_j}$ is the operator which performs the starlet transform, hard thresholds the starlet coefficients, and reconstructs an image from the thresholded coefficients. Here, $\lambda_j$ is the threshold that allows us to select only coefficients that are significant enough to represent the signal. Thresholds are updated at each iteration as described in the following paragraphs. Full details can be found in \citet{starck:book10}.
Pseudo-algorithm ~\ref{algo:MuSCADeT} shows the principle of this iterative scheme. 
MuSCADeT is an iterative process that alternates between a gradient step (line \ref{line:grad}) and a filtering of the components in transformed space through iterative hard thresholding (line \ref{line:hard}) . 

 \begin{algorithm}
 \begin{algorithmic}[1]
 \Procedure{MuSCADeT}{$Y,K,A,J,N_{iter}$}
 \State $\tilde{X} \gets 0$
 \For{$0 < i \leq N_{iter}$}
  \State $R = \mu(A^T(Y-A\tilde{X}))$
  \State {\bf Update $\lambda$}
 \State $\tilde{X} \gets \tilde{X} + R $ \label{line:grad}%{\color{gray} \% Gradient descent }
 \State $\lambda \gets mi(\frac{\lambda-K}{N_{iter}-i-6} ,MOM(X))$ \label{line:mom}
  \For{$0 < j \leq  J$} 
  \State $\tilde{x}_j \gets \Delta_{\Psi \lambda_j}(\tilde{x}_j)$ \label{line:hard}%{\color{gray} \% Hard thresholding }
 % \State $\alpha_j \gets  \Psi^t \tilde{x}_j   $ {\color{gray} \% Apply a WT on each source $x_j$ }
  %\State $\tilde{\alpha}_j  \gets Th_{\lambda}(\alpha_j)$ {\color{gray} \% soft thresholding}
 %\State $\tilde{X}  \gets  \Psi \tilde{\alpha}_j$ {\color{gray} \% wavelet reconstruction}
 \EndFor
  \EndFor
 \State \textbf{return} $\tilde{X}$ 
 \EndProcedure
 \end{algorithmic}
 \caption{MuSCADeT algorithm}
 \label{algo:MuSCADeT}
\end{algorithm}

\subsection{Thresholding strategy}
 
	Thresholding aims at selecting the coefficients in transformed space that allow us to reconstruct the desired signal only. In this case, this means that for a given component, we want to select coefficients above noise level that accounts for this component and not for the others. It is therefore crucial to devise an adequate method to adapt thresholds at each iteration of algorithm \ref{algo:MuSCADeT}.
	
	Since each iteration moves our solution for components closer to a good separation, the thresholds have to be decreased to capture fainter and fainter structures. A classical way is to operate a linear decrease for instance, where values for $\lambda_j$ are linearly sampled between an initial threshold chosen high above noise levels and a sensitivity value $K$. In general, $K$ is chosen between three and five. The sensitivity value three allows for good completeness of detected coefficients and five ensures a selection that is free from noise-related coefficients. Noise levels are computed using median absolute deviation \citep{Donoho1994}. Although linear or exponential laws are well suited to such problems \citep{Starck2004}, we choose here to rely on a more adaptive strategy based on minimum of maximums \citep[MOM, see][]{Bobin2007}: 
	
	At each iteration, we simply estimate the maximum coefficient of each component in its own morpho-spectral dictionary and choose the smallest maxima plus a margin as a threshold. If the result is smaller than the threshold given by a linear decrease, the threshold is updated with the value estimated from MOM as illustrated in line~\ref{line:mom} of the {\tt MuSCADeT} algorithm (\ref{algo:MuSCADeT}).
Full details on this thresholding scheme can be found in \citet{starck:book10}.

%\begin{eqnarray}
%&&\lambda_{min} = \underset{l}{min(max(\Psi^*x_l))} \nonumber \\
%&&\lambda_{max} = \underset{l}{max(max(\Psi^*x_l))} \nonumber \\
%&&\lambda_j = \lambda_{min} + \frac{\lambda_{max}-\lambda_{min}}{10} \label{eq:MOM}
%\end{eqnarray}	

%\begin{algorithm}
% \begin{algorithmic}[1]
% \Procedure{$\Delta$}{$x,\lambda,\Psi,N_{iter}$}
% \State $\alpha \gets \Psi^* x$
% \State $\tilde{x} \gets 0$
% \State $M_r \gets 0${\color{gray} \% Multi-resolution support}
% \State $M_r[${\bf where} $\alpha > \lambda] \gets 1$ 
% \For{$0 < i \leq N_{iter}$}
%  \State $R \gets x-\tilde{x}$
% \State $\alpha \gets \Psi^* R $ 
%\State $\tilde{x} \gets \tilde{x} + \Psi(M_r\odot \alpha )$
% \EndFor
% \State return} $\tilde{x}$ 
% \EndProcedure
% \end{algorithmic}
% \caption{Thresholding procedure}
% \label{algo:HT}
%\end{algorithm}

%=======================

%
\begin{figure*}[t!]
%\centering
\begin{tabular}{cccc}
\hspace*{-4mm}
\includegraphics[scale=0.28,trim = 4cm 0 4cm 0cm, clip = true]{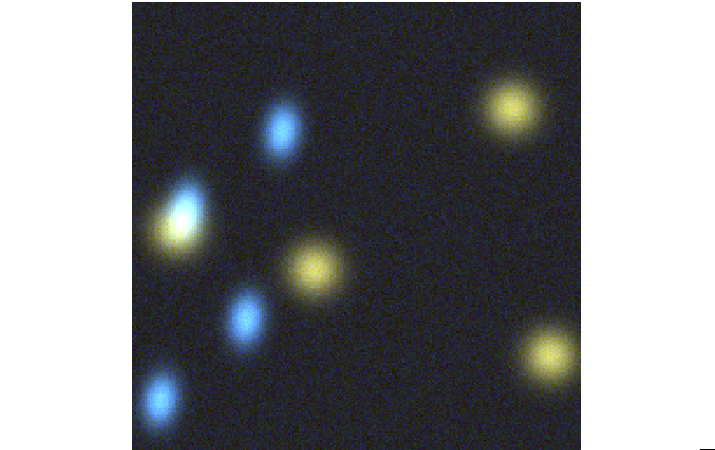} & 
\hskip -18pt
\includegraphics[scale=0.28,trim = 4cm 0 4cm 0cm, clip = true]{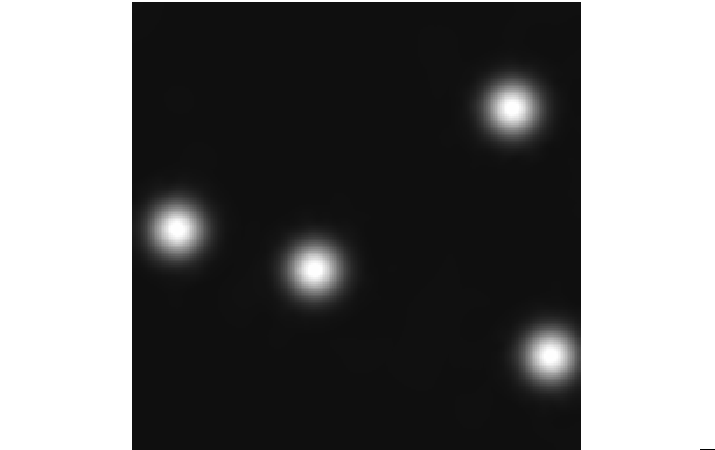} &
\hskip -18pt
\includegraphics[scale=0.28,trim = 4cm 0 4cm 0cm, clip = true]{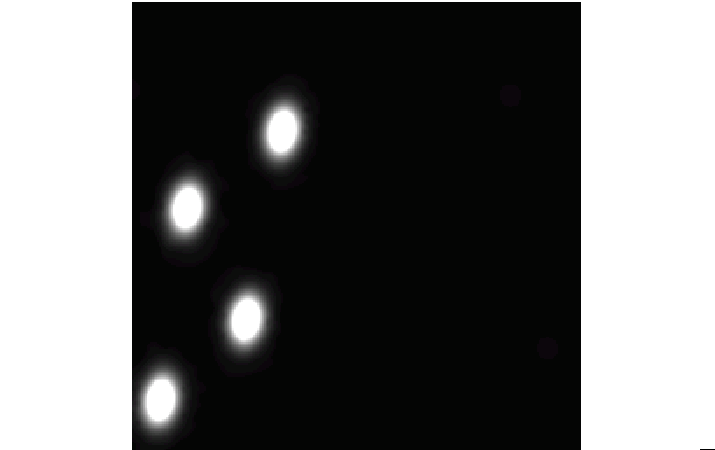}&
\hskip -18pt
\includegraphics[scale=0.28,trim = 4cm 0 4cm 0cm, clip = true]{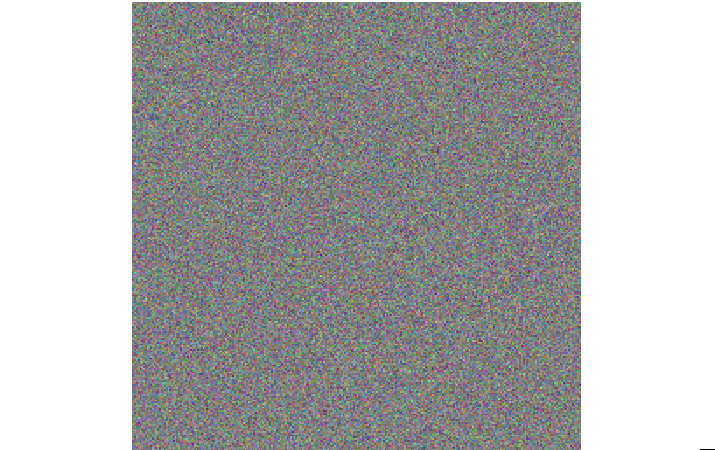} 
\end{tabular}
\caption{Source separation with {\tt MuSCADeT} in the case of a simple colour separation with SEDs estimated from PCA.  {\it From left to right}: original simulated image of colour sources, first and second components (elements of $X$ in eq.~\ref{eq:Source}) as extracted by {\tt MuSCADeT}, and the residual image after subtracting both components from the original image. \label{fig:simple}}
\end{figure*}
\begin{figure*}[t!]
%\centering
\begin{tabular}{cccc}
\hspace*{-4mm}
\includegraphics[scale=0.28,trim = 4cm 0 4cm 0cm, clip = true]{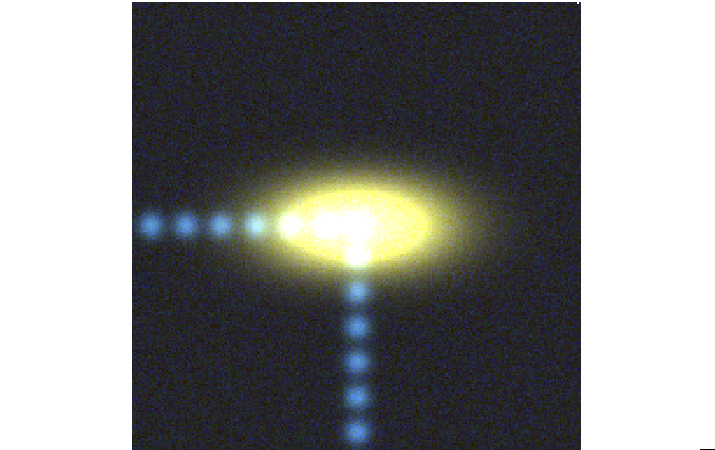} & 
\hskip -18pt
\includegraphics[scale=0.28,trim = 4cm 0 4cm 0cm, clip = true]{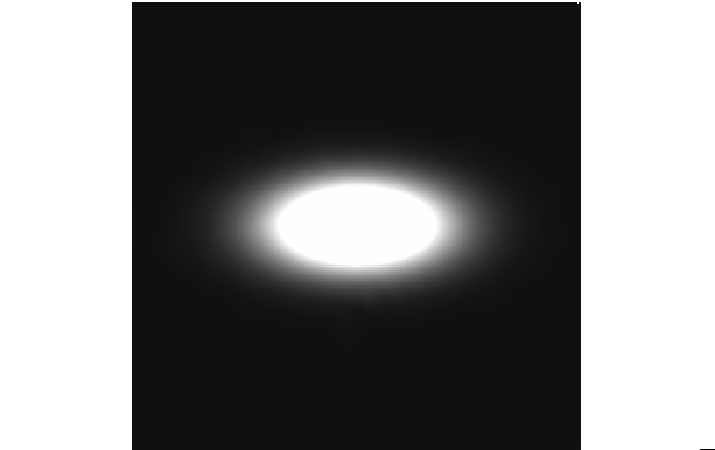} &
\hskip -18pt
\includegraphics[scale=0.28,trim = 4cm 0 4cm 0cm, clip = true]{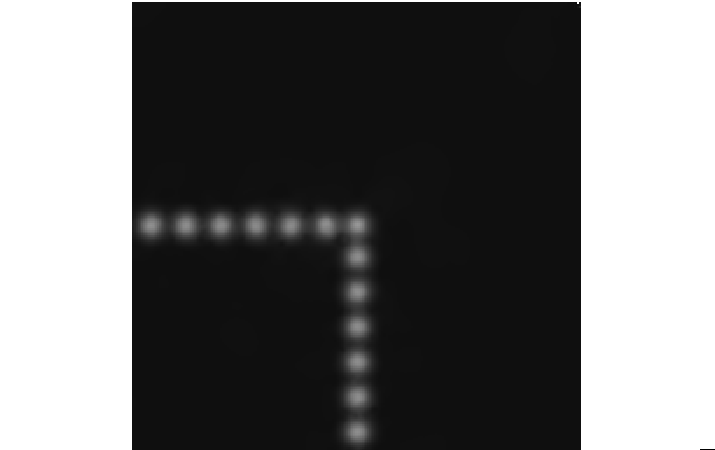}&
\hskip -18pt
\includegraphics[scale=0.28,trim = 4cm 0 4cm 0cm, clip = true]{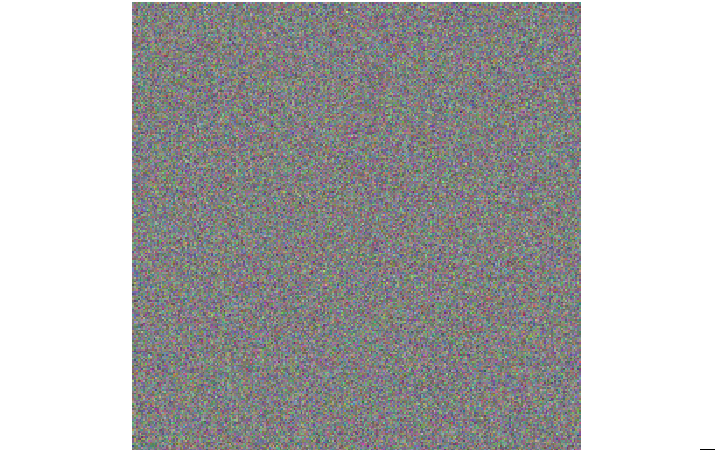}
\end{tabular}
\caption{Separation of blended sources with SEDs estimated from PCA (as in Fig.~\ref{fig:simple}).\label{fig:big}}
\end{figure*}
\begin{figure*}[t!]
%\centering
\begin{tabular}{ccc}
\hspace*{-2mm}
\includegraphics[scale=0.25]{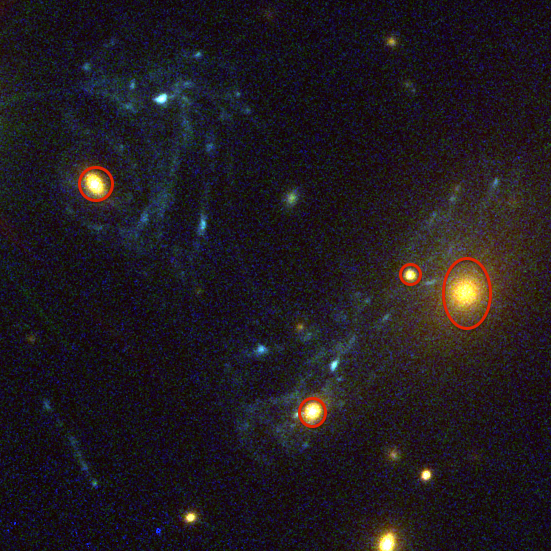}&
\hskip -10pt 
\includegraphics[width = 6.8cm, height = 5.cm]{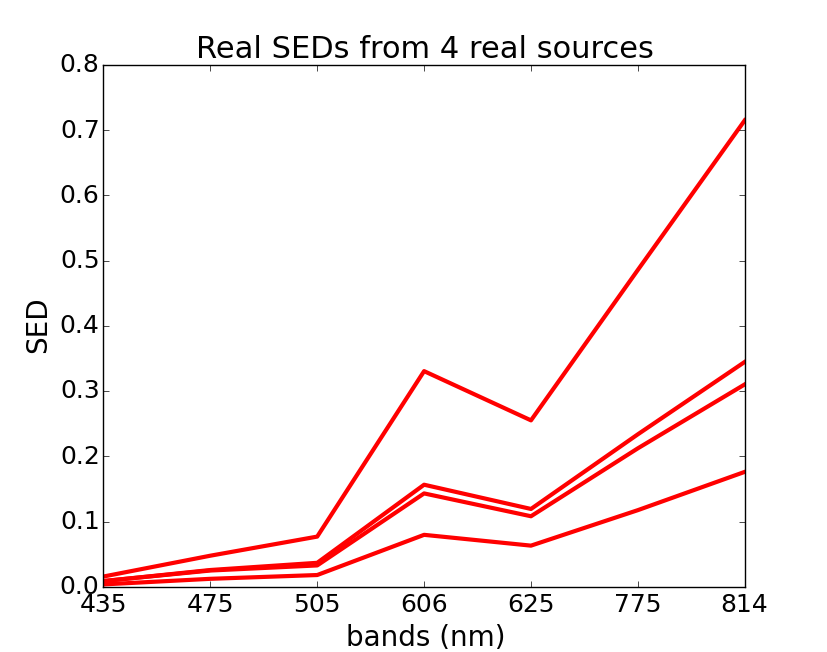}&
\hskip -15pt
\includegraphics[width = 6.8cm, height = 5.cm]{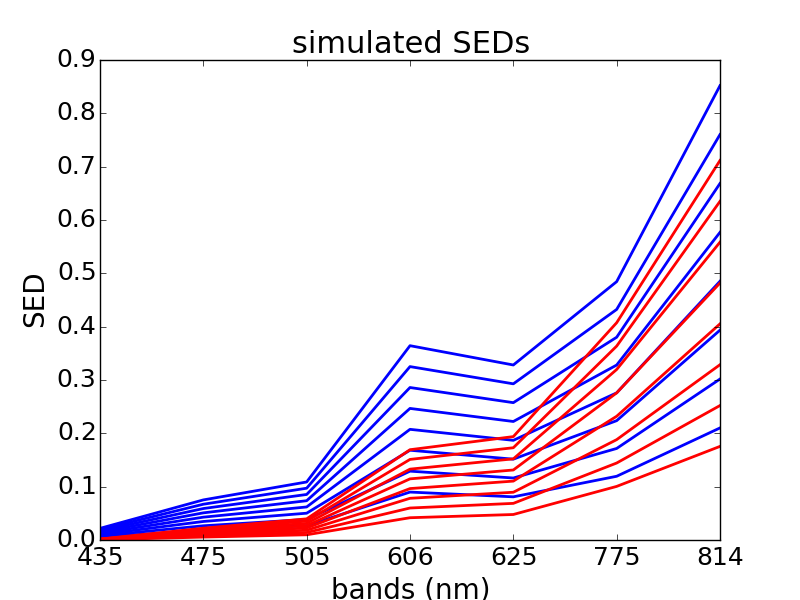}
\end{tabular}
\caption{{\it Left:} part of an HST image of the galaxy cluster MACS~J1149+2223, where the four objects used to extract SEDs are indicated with red contours. {\it Middle:} extracted SEDs. Each curve corresponds to the SEDs of the galaxies circled in red in the first panel. {\it Right:} simulated SEDs. The seven red SEDs are used to produce the upper row of galaxies in Fig.~\ref{fig:real}. The blue SEDs correspond to the lower row of galaxies (see text).} \label{fig:slopes}
\end{figure*}
\begin{figure*}[t!]
%\centering
\begin{tabular}{cccc}
\hspace*{-4mm}
\includegraphics[scale=0.28,trim = 4cm 0 4cm 0cm, clip = true]{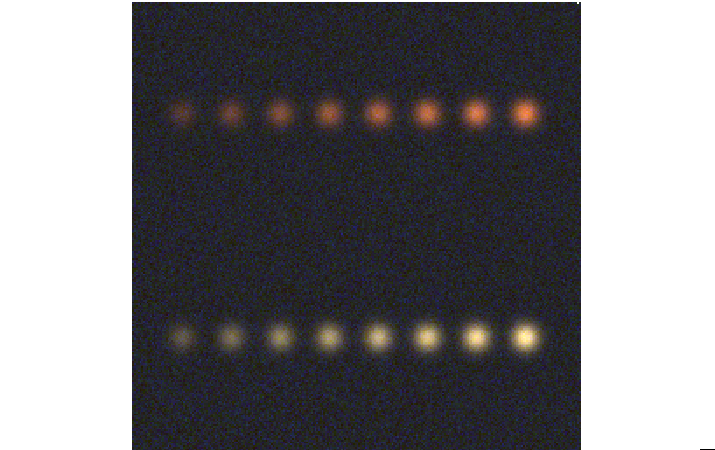} & 
\hskip -18pt
\includegraphics[scale=0.28,trim = 4cm 0 4cm 0cm, clip = true]{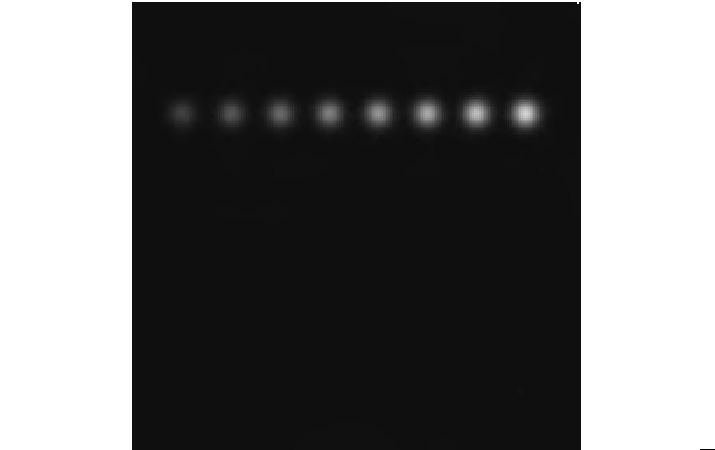} &
\hskip -18pt
\includegraphics[scale=0.28,trim = 4cm 0 4cm 0cm, clip = true]{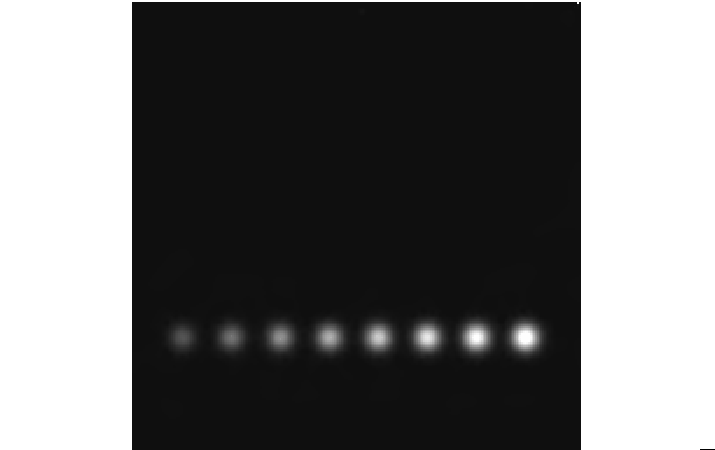}&
\hskip -18pt
\includegraphics[scale=0.28,trim = 4cm 0 4cm 0cm,  clip = true]{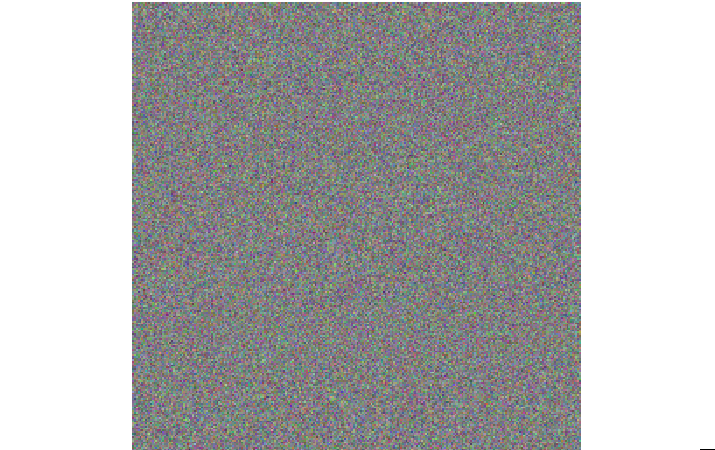}
\end{tabular}
\caption{Results of a separation of sources with poorly known SEDs (as in Fig.~\ref{fig:simple}).\label{fig:real}}
\end{figure*}

\section{Tests on simulations}
\label{sec:Test}

In this section, we present several tests conducted on simulated multi-band images that emulate realistic problems. We show that {\tt MuSCADeT} is able to separate highly blended sources and is robust to the approximation made that several astronomical sources (e.g. galaxies in a cluster) with similar but, not identical, SEDs can be considered a single source.

\subsection{Simulations}
\label{sub:Toy}

In all following simulations we generate objects with Gaussian profiles and realistic SEDs. For the first two simulations, all SEDs were extracted from real HST observations of the galaxy cluster MACS~J1149+2223 cluster \citep{Kelly2015} using the PCA method described in Sect.~\ref{sec:PCA}. Each simulation comprises seven bands. We also add a white Gaussian noise with standard deviation $\sigma=0.01$. An example of the impact of higher noise levels on our image separation is given in appendix \ref{ap:SNR}.

	We first apply {\tt MuSCADeT} to data with four red objects with exactly the same SED and four blue objects with exactly the same SED. All objects have elliptical Gaussian profiles. Three of the simulated bands are used to produce the colour image in Fig.~\ref{fig:simple}. Results of the separation after 100 iterations of {\tt MuSCADeT} are shown in the other panels of fig \ref{fig:simple}. We observe no contamination between sources since no structure contaminates either component. Since the residuals present no structure, we also conclude that each component has been fully reconstructed. 

	The second simulation tests the deblending capacities of our algorithm in the ideal case where there is no SED variation between objects of a same component. We generate an extended elliptical Gaussian profile at the centre of the image affected with a red SED. Thirteen blue profiles are evenly spread between the centre and edges of the image such that profiles at the centre are completely blended. Results of the separation are shown in Fig.~\ref{fig:big} along with a colour image of our simulated objects. No contamination between sources is visible either and residuals show no structure. We see in particular no difference between blue sources at the centre of the image (the most blended ones) and blue sources on the edges (less blended), meaning that the deblending of each profile is successful whether a profile is highly blended or not. We used 200 iterations of {\tt MuSCADeT} to produce this result.

	The last simulation tests the robustness of our algorithm to SED variation across objects from a same component. To account for realistic SED variation, we extracted the SEDs from four red galaxies in the MACS J1149+2223 cluster (see \ref{fig:slopes}), which appear to have the same colour by integrating the flux in each galaxy profile. The resulting SEDs can be seen in the middle panel of fig \ref{fig:slopes}. We recorded the slopes of the SEDs and generated a set of eight slopes linearly spread between the maximum and minimum slope estimated. These slopes are then applied to SEDs extracted from cluster MACS J1149+2223 via PCA (see Fig.~\ref{fig:slopes}). This way, we have two sets of SEDs that account for red and blue sources and that mimic a range of variations as observed in real images. Sixteen Gaussian profiles (eight red, eight blue) are then generated and each of them is associated with one of the previously generated SEDs. The left panel of Fig.~\ref{fig:real} shows three bands of the simulated images as RGB images. Figure~\ref{fig:real} shows a colour image of our simulated images and the result of a separation by MuSCADeT. We see again that no structure appears in the residuals and no contamination is found between components. However, the great similarity between SEDs from different components forced us to increase the number of iterations of {\tt MuSCADeT} to 5000 in order to obtain such results, thus increasing computation time. In each, we used the full algorithm described in this paper including an automated estimation of SEDs through PCA.

\section{Application to real data}
\label{sec:Data}

 \subsection{Lens-source separation on MACS~J1149+2223}
We now apply {\tt MuSCADeT} to real multi-band data and we compare the performances to traditional model fitting. 

We use the deep HST data set of the galaxy cluster MACS~J1149+2223 \citep["Refsdal SN";][]{Kelly2015} to carry out this experiment, we show bright cluster members producing strongly lensed images of a distant spiral galaxies with clumpy structures. Our goal is to separate the data into two sources containing the foreground lens galaxies and background lensed object(s). As the cluster contains many member galaxies and as the background galaxy has complex structure, the deblending task is challenging, making this data set a good test for our method.

MACS~J1149+2223 has been observed with the ACS in seven bands: F435W, F475W, F555W, F606W, F625W, F775W and F814W (proposal ID: 12068, principal investigator: M. Postman), providing a good spectral coverage for {\tt MuSCADeT} to work. The data are publicly available from the STScI website\footnote{\url{https://archive.stsci.edu/}} and drizzled so that the combined frames in each band have the same orientation and pixel scale. 

\begin{figure*}
\centering
\begin{tabular}{cc}
\includegraphics[width = 7.7cm, height = 7.7cm,clip = true]{Figures/All_Refsdal_2000.png} & 
\hskip -10pt
\includegraphics[width = 7.7cm, height = 7.7cm, clip = true]{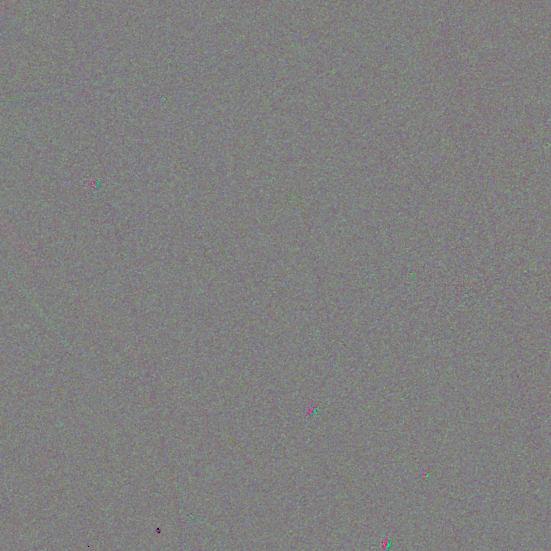}\\
\includegraphics[width = 7.7cm, height = 7.7cm,clip = true]{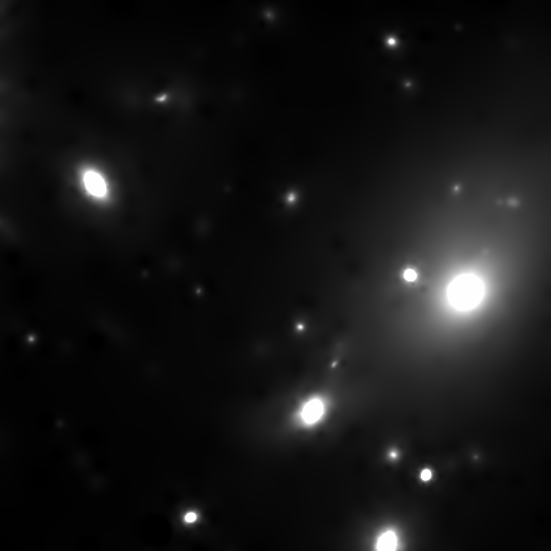} &
\hskip -10pt
\includegraphics[width = 7.7cm, height = 7.7cm, clip = true]{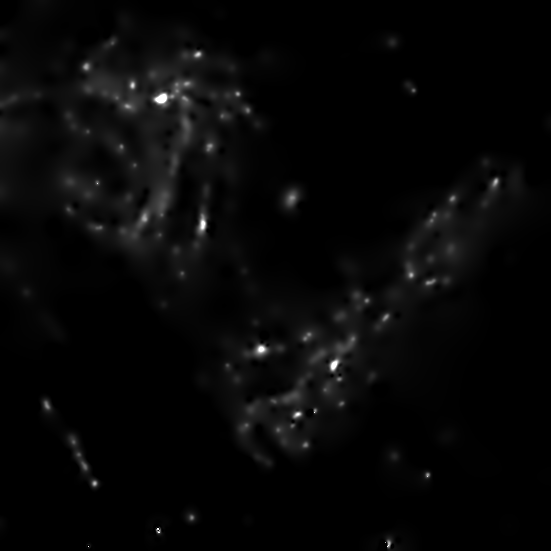}\\
\includegraphics[width = 7.7cm, height = 7.7cm,clip = true]{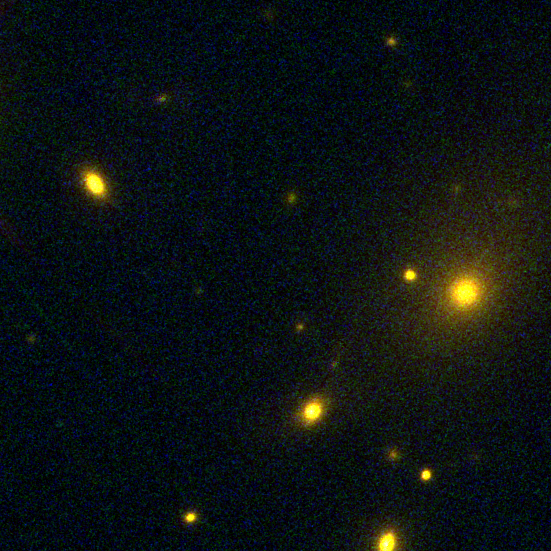} &
\hskip -10pt
\includegraphics[width = 7.7cm, height = 7.7cm, clip = true]{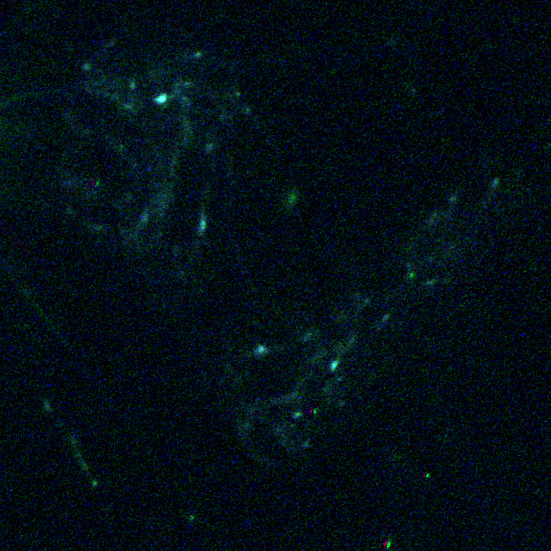}
\end{tabular}
\caption{Application of {\tt MuSCADeT} to the MACSJ1149+2223 cluster.  {\it Top left:} colour image generated using the F475W, F606W, and F814W bands. Our goal is to separate the foreground cluster galaxies (red) from the background lensed galaxy (blue). {\it Middle:} the two sources extracted with 2000 iterations of {\tt MuSCADeT}. {\it Bottom left:} original image minus the blue source found by {\tt MuSCADeT}. {\it Bottom right:} same as the bottom left, with the red source subtracted. {\it Top right:} residual image obtained after subtraction of the two estimated sources from the original data. \label{fig:refsmall}}
\end{figure*}

We estimate both SEDs in the mixing matrix using our PCA technique. The result after 2000 iterations of {\tt MuSCADeT} is shown in Fig.~\ref{fig:refsmall}, where the two separated sources are shown in the middle panels. In the two lower panels of the figure, we also show the result of the subtraction of source 1 and source 2 from the original data. The colour scale in these images is the same as in the original data. The overall residual image, i.e. with both sources subtracted from the data, is shown with $\pm 5\sigma$ cut levels. 

The source separation works very well, with the exception of a few objects with "green" SEDs. One such object is visible in the centre of the image, resulting in a signal in both red and blue sources. This is an intrinsic limitation to our algorithm that separates objects using a limited number of sources, each one with its own SED. Although the SEDs do not need to be known perfectly (as shown in our tests with simulated images), objects with SEDs falling "in between" the SEDs allocated to each source, may lead to inaccurate separation. A possible mitigation strategy is to add extra sources to the decomposition with for example a blue, a green, and a red SED. Deciding whether to do this or not depends on the exact scientific application. 

In our example, we do not take the PSF convolution into account. A variation of the PSF with wavelength can introduce artefacts in the separation, especially for objects whose angular size is comparable with the size of the PSF. The central parts of galaxies show such structures and, indeed, in the present data small structures are seen in the residual image at the location of the foreground galaxies. Introducing the PSF convolution can be done in principle, and it might be needed at least for some applications, but at a cost of increased computation time and complexity in the minimisation process. 

%We find that the centre of the lensed galaxy is partially present in the component associated to red galaxies (top left corner). Indeed, we can see on infra-red images \citep[see][]{Kelly2015} that the centre of this galaxy is intrinsically red.

\subsubsection{Comparison with profile fitting}

\begin{figure*}
%\centering
\begin{tabular}{ccc}
\hspace*{-2mm}
\includegraphics[scale=0.31]{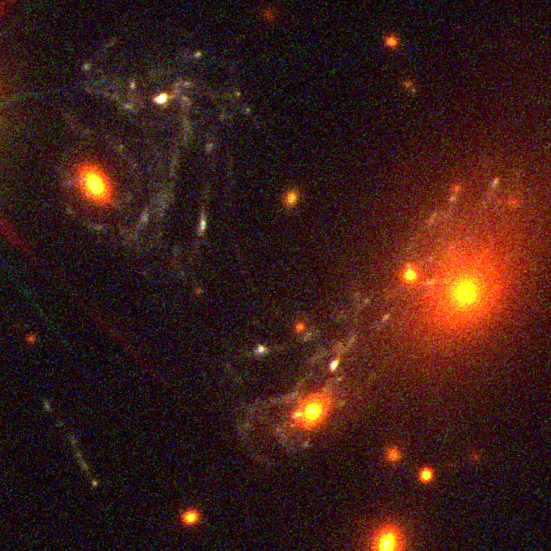}&
\hskip -10pt
\scalebox{-1}[1]{\includegraphics[scale=0.31, clip = true, angle = 90]{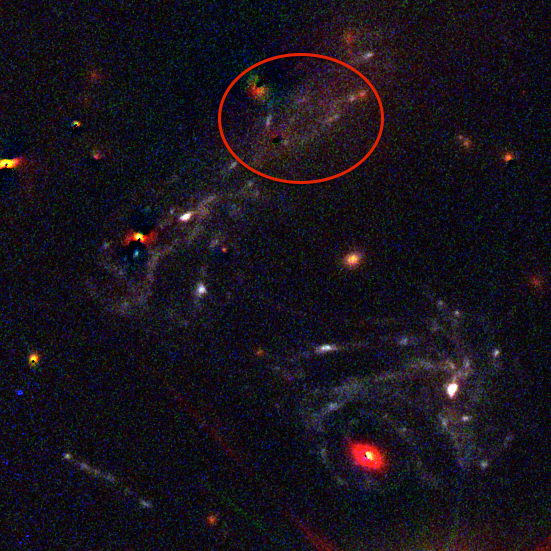}}&
\hskip -10pt
\includegraphics[scale=0.31]{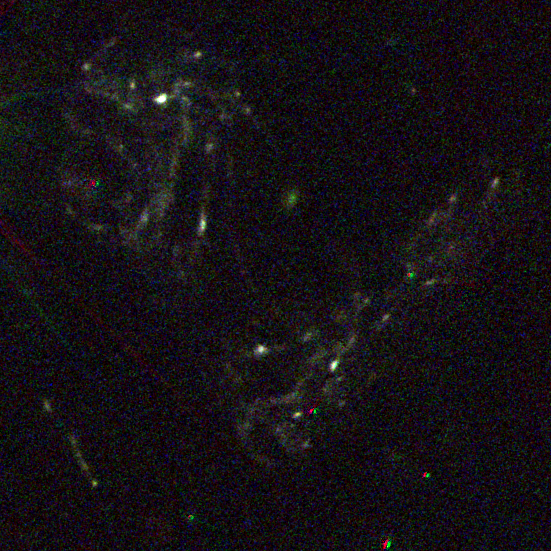}
\end{tabular}
\caption{Comparison between the {\tt MuSCADeT} and that {\tt galfit} separations. The original colour image of MACS~J1149+2223  is shown on the left, followed by the {\tt galfit} subtraction of the galaxy members and the subtraction using {\tt MuSCADeT}. The colour cuts are exactly the same in all three images. The red ellipse in the middle panel indicates an area where the {\tt galfit} fit leads to under-subtraction of a galaxy halo.} \label{fig:galfit}
\end{figure*}

A popular way to carry out source separation in galaxy clusters is to fit two-dimensional profiles to the cluster members and to subtract from the data. We apply such a procedure to the HST data of the MACS~J1149+2223 cluster and compare with {\tt MuSCADeT}. In this exercise, we identified nine red objects with elliptical profiles that we fit using the {\tt galfit} \citep{Peng2002} software. Each red object is fitted with either one or two Sersic profiles depending on its morphology. Fitting is performed in each of the three bands used to build the colour image (F475W, F606W, and F814W) using the same initial conditions. Since no PSF was used in applying {\tt MuSCADeT} to this cluster, we used a Dirac function as an input PSF for {\tt galfit}. The result is presented in the middle panel of Fig.~\ref{fig:galfit} along with the {\tt MuSCADeT} result. We note that: 

%It is also important to notice that {\tt MuSCADeT} is able to capture complex features from the lensed galaxy, which current model fitting method fail to achieve.

%Although {\tt galfit} subtracts most of red galaxies profiles, the method presents several major drawbacks compared to our {\tt MuSCADeT} algorithm:
	
\begin{itemize}

\item Profile fitting leaves significant artefacts in the central parts of the fitted galaxies even when a double Sersic model is used.

\item Profile fitting does not model the extended halos of galaxies well, as shown in the circled region in middle panel of Fig.~\ref{fig:galfit}. This is both because extended halos are not simple analytical profiles and because the many background lensed structures influence the fit. In principle, these structures can be masked but 1- designing the mask can be difficult and is time consuming, 2- in some parts of the image there is no clean area with no contamination by the background object. The mask would take the majority of the data.

\item The top left lensed image appears to be more blue after {\tt galfit}'s run than with {\tt MuSCADeT}. This could mean that {\tt MuSCADeT} performs badly at extracting extended halo. However, it would imply that too much signal from the red source has been attributed to the blue source. Therefore, when subtracting the blue source from the original images one would see holes in the extended profiles of the red galaxies that are not observed here (see lower left panel of Fig.~\ref{fig:refsmall}). This implies that {\tt galfit} is overfitting the extend halo to compensate for blue structures, as pointed out in the previous note, thus removing part of the flux from blue sources.

\item The human time involved in profile fitting can be a limiting factor for large data sets. The user has to decide where to put a galaxy and to find a strategy to estimate the initial guesses for the many parameters involved in the fit. {\tt MuSCADeT} is fully automated procedure with only one parameter to be chosen, i.e. the sensitivity value $K$ involved in the threshoding scheme.

\end{itemize}

\subsection{Bulge-disk separation on spiral galaxy in Abell~2744}

Another possible application of our algorithm is the separation of coloured disk and bulge components in a spiral galaxy. To illustrate this, we use a spiral galaxy in the galaxy cluster Abell~2744, which was imaged with the HST as part of the Hubble Frontier Fields programme. 

\begin{figure*}[t!]
%\centering
\begin{tabular}{cccc}
\hspace*{-4mm}
\includegraphics[scale=0.28,trim = 4cm 0 4cm 0cm, clip = true]{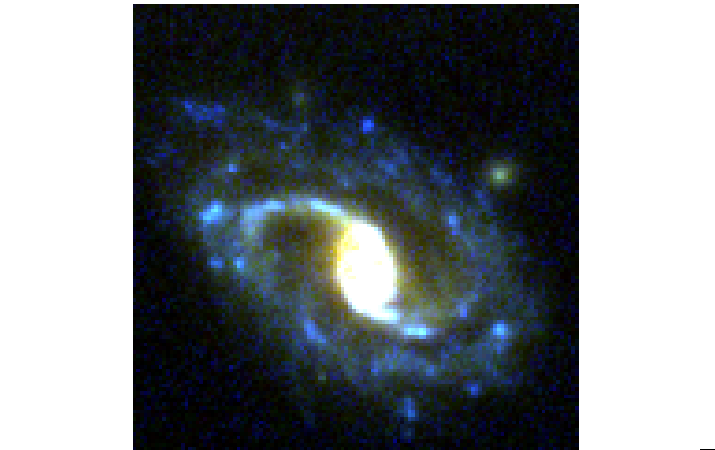} & 
\hskip -18pt
\includegraphics[scale=0.28,trim = 4cm 0 4cm 0cm, clip = true]{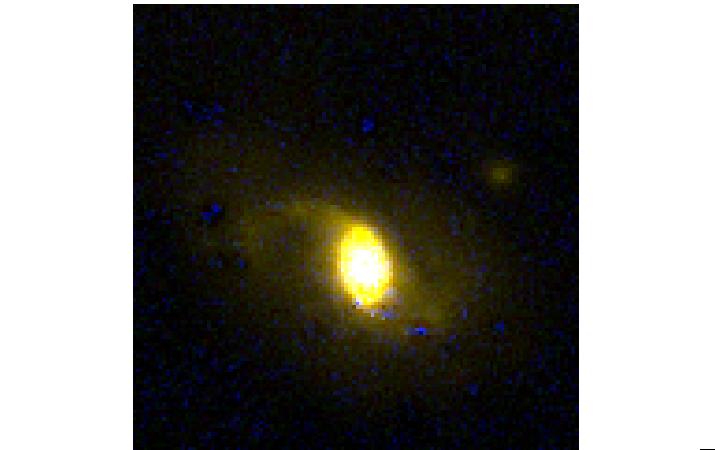} &
\hskip -18pt
\includegraphics[scale=0.28,trim = 4cm 0 4cm 0cm, clip = true]{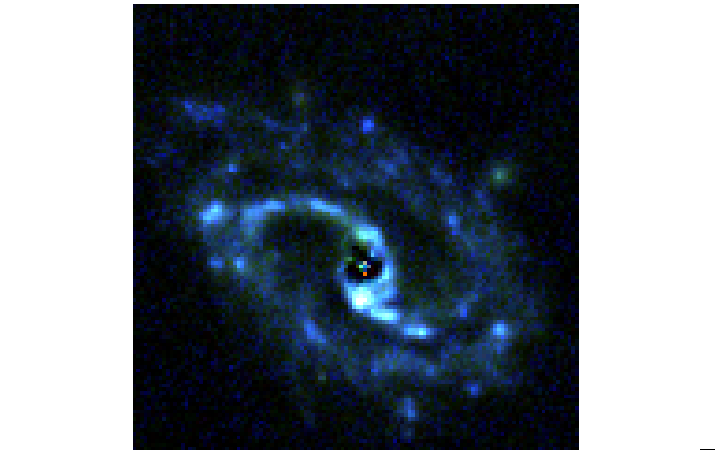}&
\hskip -18pt
\includegraphics[scale=0.28,trim = 4cm 0 4cm 0cm,  clip = true]{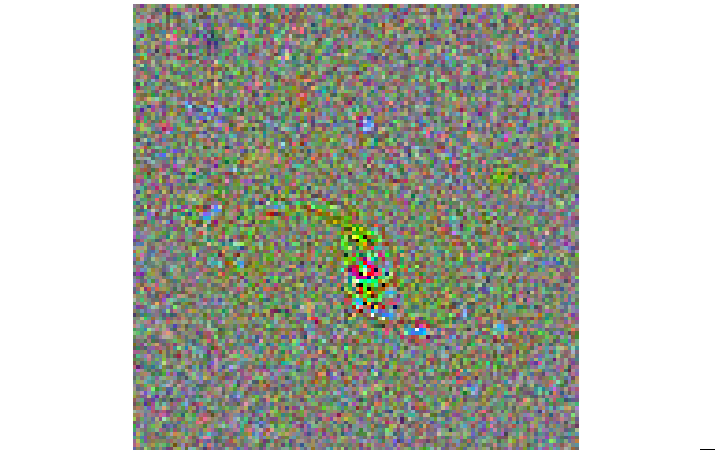}
\end{tabular}
\caption{ Separation of the red and blue stellar populations in a spiral galaxy. From left to right we show the original HST colour image of a spiral galaxy in the galaxy cluster Abell~2744, the same colour image after subtraction of the blue component estimated with {\tt MuSCADeT}, the colour image after subtraction of the red component estimated with {\it MuSCADeT}, and the residual image after subtraction of both components. The three first images have the same colour cuts. The residuals are shown with cut levels set to five times the noise level in each colour channel.\label{fig:2744}}
\end{figure*}

Abell~2744 was observed with the ACS in three bands: F435W, F606W, and F814W, resulting in the colour image in Fig.~\ref{fig:2744}(Proposal ID: 11689, principal investigator: R. Dupke). The data are publicly available from the STScI  website\footnote{\url{https://archive.stsci.edu/pub/hlsp/frontier/}} and drizzled so that the combined frames in each band have the same orientation and pixel scale. 

Spiral galaxies represent an obvious test bench for our algorithm as they are composed of a red bulge of older stars and of a blue disk dominated by young stars. The ability to separate both colour components in an unbiased way allows us to trace stellar populations in galaxies and/or highlight morphological relations between bulge and disk shapes. Numerous examples can be found in the literature of studies where bulge and disk components are separated morphologically, using profile-fitting techniques featuring Sersic, exponential, or DeVaucouleurs functions in either single or multiple bands \citep[e.g.][]{Vika2014, Maltby2012, Pastrav2013, Kendall2011}. 

With {\tt MusCADeT}, we are able to capture the morphology of the young and old stellar populations of the spiral galaxy in Abell~2744 under the single assumption that they do not display the same colour. Our decomposition in Fig.~\ref{fig:2744} shows an elliptical red bulge with extended and smooth features along the spiral arms, which elliptical profile-fitting methods would fail to model. The separation of the blended blue and red components in Fig.~\ref{fig:2744} is overall excellent except for slight cross-contamination of the two colour channels due to features smaller than the PSF size. Adding the PSF in our separation technique is under development. However, even without this refinement, the decomposition presented in our example would be impossible to achieve with profile-fitting techniques given the morphological complexity of the galaxy.

%Indeed, the small point-like features at the centre of the middle right panel and in the spiral arms in the middle left panels as well as in the residuals are attributed to contamination from varying PSF. Since in this case we have only three bands, spectrally farther apart from on another than in the case of MACS~J1149+2223, we are much more sensitive to the PSF variation between bands (pas tres important ici, non ?). 

\section{Conclusion}

We have developed a new model-independent tool to deblend the images of astronomical objects based on their colour contrast. This is made possible via modern techniques for image analysis based on sparsity of objects in an appropriate dictionary. 
%This makes the separation an automated method free of any light model for the sources to separate. 

More specifically, we created a morpho-spectral dictionary combining Starlets and SEDs to separate images of galaxies with different colours. We show from simulated data that our algorithm is robust against very strong blending as well as against SED variations among objects belonging to the same colour component (source) of our model. SED variations across the field result in an increase of the computing time by one order of magnitude with respect to the ideal case where all objects in a given source have the same SED. This does not hamper a reliable source separation, however.

The method is successfully applied to the deep HST images of the galaxy cluster MACS~J1149+2223, in which we separate the foreground red galaxies from background blue clumpy lensed galaxies, and to Abell~2744, where we separate the red and blue stellar populations of a spiral galaxy. This is done in an automated way and with better efficiency that with standard profile fitting. All codes used to produce the results presented here are made freely available for the sake of reproducible research and to make our tool usable for the community.

Future developments of our method include accounting for the PSF in each band and including explicit  SED variations across the field of view. These SED variations multiply the complexity of the problem  by $N_p\times N_s\times N_b$, but the effect of the increased complexity can likely be minimised by using sparse priors and physical constrains on the SED profiles. Also, the increased computation time resulting from the extra complexity should be partially compensated by a reduced number of iterations. 

The deblending method described here was devised speci\-fically to address the problem of object deblending in the case of strong lensing systems. Deblending is essential in this case to see small and faint details in the lensed structures, which are free of contamination by the bright foreground. The range of applications of the method is nevertheless much broader. Among the numerous possible applications are the identification of star-forming regions in galaxies, model-independent bulge-disk decompositions of galaxies, or even the improvement of photometric redshift in large sky surveys where blending can be a serious issue given the depth of the data.

\begin{acknowledgements} 
This work is supported by the Swiss National Science Foundation (SNSF), and by the European Community through grant PHySIS  (contract no. 640174) and  DEDALE  (contract no. 665044) within the H2020 Framework Program. The HST images used in this paper were obtained from the Mikulski Archive for Space Telescopes (MAST). STScI is operated by the Association of Universities for Research in Astronomy, Inc., under NASA contract NAS5-26555. Support for MAST for non-HST data is provided by the NASA Office of Space Science via grant NNX09AF08G and by other grants and contracts. We would like to thank the members of the CosmoStat team for useful discussions on inverse problem solving using sparse prior.
\end{acknowledgements}

\bibliographystyle{aa} % -> aa.bst
\bibliography{biblio} % -> papers.bib

%%%%%%%%%%%%%%%%%%%%%%%%%%%%%%%%%%%%%%%%%%%%%%%%%%%%%%%%%%%%%%%%%%%%%

\begin{appendix}

\section{Reproducible research}
\label{sec:repro}
In the spirit of carrying out reproducible research, we make pubic all codes and resulting products describes in this paper. Table~\ref{public_code} lists all products that will be made available along with this paper. The {\tt MuSCADeT} itself is made available as well as input files and routines needed for all benchmark tests and for the application to real data. The routines provide simple examples of how to execute the {\tt MuSCADeT} algorithm. We encourage potential users to modify them as they wish for their own scientific applications. 
\begin{table*}[h!]
  \centering
  \begin{tabular}{@{} lcl @{}} % Column formatting, @{} suppresses leading/trailing space
  Product name  &Type & Description\\
  \hline
  Software products:\\
  {\tt MuSCADeT}  & python package  & includes {\tt MuSCADeT} implementation and visualisation tools  \\
  Routines: \\
  {\tt Example\_simple.py}  & code (python)  & routines to reproduce Fig.~\ref{fig:simple}.\\
  {\tt Example\_big.py}  & code (python)  & routines to reproduce Fig.~\ref{fig:big}.\\
  {\tt Example\_real.py}  & code (python)  & routines to reproduce Fig.~\ref{fig:real}.\\
  {\tt Example\_refsdal.py}  & code (python)  & routines to reproduce Fig.~\ref{fig:refsmall}.\\
  {\tt Example\_2744.py}  & code (python)  & routines to reproduce Fig.~\ref{fig:2744}.\\
  {\tt Example\_SNR.py}  & code (python)  & routines to reproduce Fig.~\ref{fig:SNR}.\\
  {\tt Example\_nottoosimple.py}  & code (python)  & An other example of a {\tt MuSCADeT} run on simulations.\\
  Simulations:\\ 
  {\tt Cube.fits} & fits data cube & cube with all simulated images for each benchmark \\
  {\tt Simu\_A.fits} & fits table & table with the simulated spectra used in our simulations \\
  \hline

  \end{tabular}
  \caption{List of products made available in this paper in the spirit of reproducible research. All above material is available here:  \url{http://lastro.epfl.ch/page-126973.html}. \label{public_code}}
  \label{tab:reproducible_ps}
\end{table*}

\section{Deblending noisy data with known SED}
\label{ap:SNR}

We show here that the {\tt MuSCADeT} algorithm is capable of separating blended sources even in the case of high noise levels. We generate simulations of blended objects as in Fig.~\ref{fig:big}, but with a noise level ten times larger. The high noise levels along with the strong blending make it hard for the PCA estimator to estimate a good SED for the separation. 
	
\begin{figure*}[h!]
%\centering
\begin{tabular}{cccc}
\hspace*{-4mm}
\includegraphics[scale=0.28,trim = 4cm 0 4cm 0cm, clip = true]{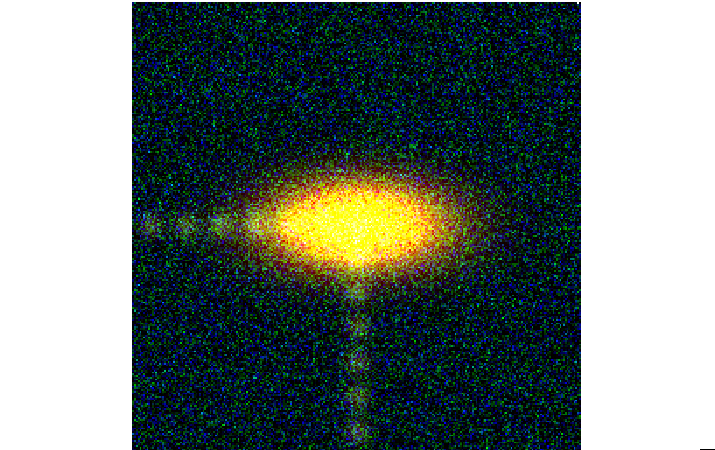} & 
\hskip -18pt
\includegraphics[scale=0.28,trim = 4cm 0 4cm 0cm, clip = true]{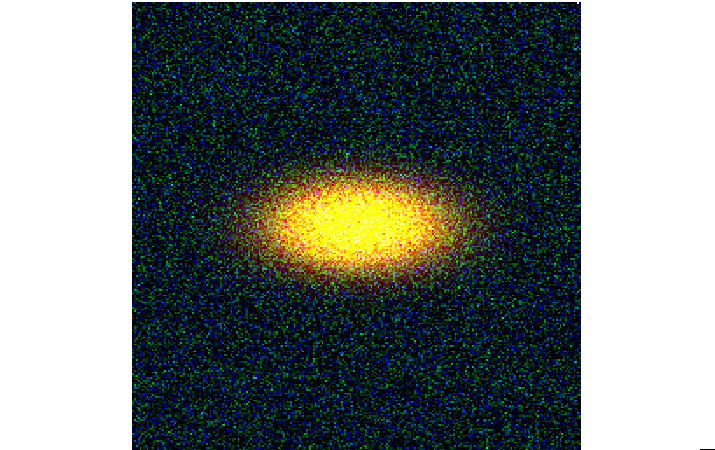} &
\hskip -18pt
\includegraphics[scale=0.28,trim = 4cm 0 4cm 0cm, clip = true]{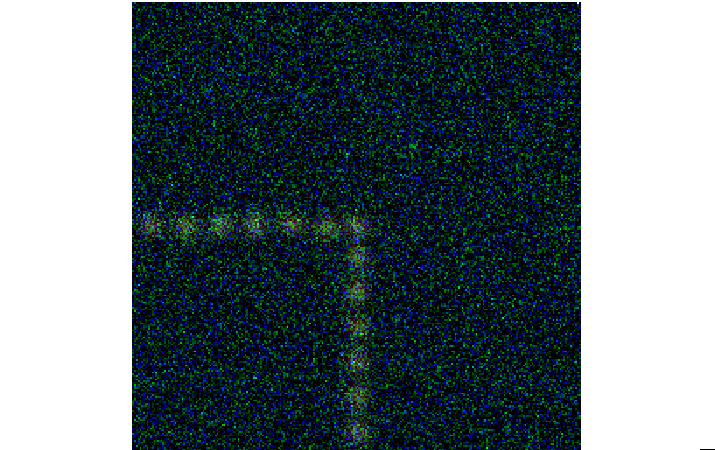}&
\hskip -18pt
\includegraphics[scale=0.28,trim = 4cm 0 4cm 0cm,  clip = true]{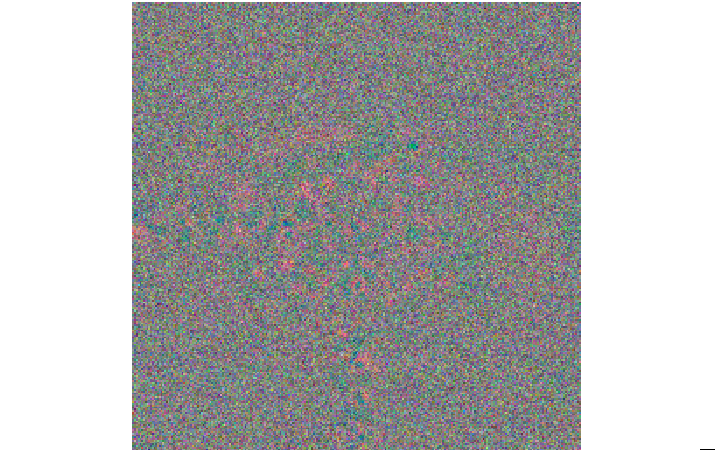}
\end{tabular}
\caption{ Separation of blended sources at low S/N. From left to right are shown the original simulated images, the original image after subtraction of the blue component as estimated from {\tt MuSCADeT}, the original image after subtraction of the red component and the residual image after subtraction of both components.\label{fig:SNR}}
\end{figure*}

Although the PCA technique fails in such conditions, the main feature of our algorithm, which is the morphological component analysis-based inversion, still manages to estimate good sources. The price to pay in this case is that the SEDs must be known. Figure ~\ref{fig:SNR} shows the result of a separation performed by {\tt MuSCADeT} on very noisy data using known SEDs, showing that our algorithm is still able to separate sources. Our PCA SED estimator might replaced in the near future to cope with noisy data. For the present example, we decided to show the residuals after separation of both colour components (middle panels of Fig.~\ref{fig:SNR}) to show that we estimate the sources down to the noise level.
	
\end{appendix}

%%%%%%%%%%%%%%%%%%%%%%%%%%%%%%%%%%%%%%%%%%%%%%%%%%%%%%%%%%%%%%%%%%%%%
\end{document}